\documentclass[aps,reprint,superscriptaddress]{revtex4-1}
\usepackage{CJK}
\usepackage{siunitx}
\usepackage{graphicx}
\usepackage{mathtools,amsthm}
\usepackage{epstopdf} 

\begin{document}

\title{Domain formation mechanism of the Si(110)``$16\times2$'' reconstruction}
\author{N. K. Lewis}
\affiliation{The Photon Science Institute, University of Manchester, Oxford Road, Manchester, M13 9PL, UK}
\affiliation{The Cockcroft Institute, Daresbury Laboratory, Sci-Tech Daresbury, Warrington, WA4 4AD, UK}
\author{N. B. Clayburn}
\affiliation{Jorgensen Hall, University of Nebraska, Lincoln, Nebraska 68588-0299, USA}
\author{E. Brunkow}
\affiliation{Jorgensen Hall, University of Nebraska, Lincoln, Nebraska 68588-0299, USA}
\author{T. J. Gay}
\affiliation{Jorgensen Hall, University of Nebraska, Lincoln, Nebraska 68588-0299, USA}
\author{Y. Lassailly}
\affiliation{Physique de la Mati{\`e}re Condens{\'e}e, CNRS-Ecole Polytechnqiue, 91128 Palaiseau C{\'e}dex, France}
\author{J. Fujii}
\affiliation{Istituto Officina dei Materiali (IOM), CNR, AREA Science Park Basovizza, I-34149 Trieste, Italy}
\author{I. Vobornik}
\affiliation{Istituto Officina dei Materiali (IOM), CNR, AREA Science Park Basovizza, I-34149 Trieste, Italy}
\author{W. R. Flavell}
\affiliation{The Photon Science Institute, University of Manchester, Oxford Road, Manchester, M13 9PL, UK}
\author{E. A. Seddon}
\affiliation{The Photon Science Institute, University of Manchester, Oxford Road, Manchester, M13 9PL, UK}
\affiliation{The Cockcroft Institute, Daresbury Laboratory, Sci-Tech Daresbury, Warrington, WA4 4AD, UK}
\date{\today}
\begin{abstract}
The main factor that determines which of the two domains form upon reconstruction of the Si(110)``$16\times2$'' surface has been investigated. LEED and STM images showed that the domain orientation was independent of the heating current direction used to induce the Si(110)``$16\times2$'' reconstruction. Reciprocal-space lattice models of the reconstruction allowed for the correct identification of the domain orientations in the LEED images and confirm that the reconstruction is 2D-chiral. It is proposed that the domain orientation upon surface reconstruction is determined by the direction of monoatomic steps present on the Si(110) plane. This is in turn determined by the direction at which the surface is polished off-axis from the (110) plane.
\end{abstract}
\maketitle

\section{Introduction}
Silicon nanowires and nanomeshes provide opportunities for creating novel nanoelectronics and optical devices \cite{Cui,Holmes,Ma,Zhong}. It has long been established that the Si(110)``$16\times2$'' reconstruction is made up of extensive domains \cite{An,Stekolnikov,Kang} thus it has been used as a template for nanowires and nanomeshes \cite{Hong}. Understanding the domain formation mechanism for this reconstruction is important in the light of the growing interest in nanowire technology \cite{Hong,Hong2015}. In addition, this reconstruction has been reported to be 2D-chiral \cite{An,Hel,Yamada}. This makes it interesting because it is a low-index surface which exhibits single enantiomers over large areas ($\si{mm^2}$) \cite{Yamada} without the adsorption of chiral molecules \cite{Lorenzo,Richardson}. Overall, a detailed understanding of how the surface reconstructs is key to reliable generation of nanowires as well as the understanding of low-index surface chirality.

The reconstruction process for the Si(110)``$16\times2$'' surface has been extensively studied at various temperatures using scanning tunneling microscopy (STM). Observations of the surface above $770\,\si{\degreeCelsius}$ have shown that it exhibits vicinal (17 15 1) and (15 17 1) surfaces \cite{Olshanetsky,Yamamoto1994,Yamamoto1994I,Ichikawa} with fluctuating steps \cite{Yamamoto2000}. Step bunching, in which fluctuating steps bunch together, occurs as the temperature is reduced below $760\,\si{\degreeCelsius}$ resulting in a facetted surface \cite{Jeong,Williams,Phaneuf}. This step bunching is a result of the nucleation of the ``$16\times2$'' reconstruction. It begins at a vicinal step and expands on the (110) terraces \cite{Yamamoto1994,Alguno,Ichikawa} such that the monoatomic steps move together. The facetted surface consists of 7 to 8 steps \cite{Yamamoto1994}, determined by the reduction in free energy forming the ``$16\times2$'' from the 1$\times$1 surface \cite{Hibino}. This has also been observed on vicinal Si(111) surfaces for the 7$\times7$ reconstruction \cite{Hibino}. 

\begin{figure}
	\centering
	\includegraphics[width=0.8\linewidth]{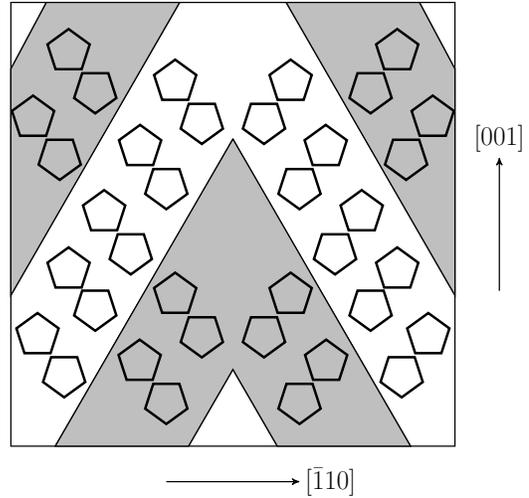}
	\caption{Schematic diagram of the double domain structure of the Si(110)``$16\times2$'' reconstruction. The gray and white segments correspond to the periodic channel structure with `pairs of pentagons' on both levels. The channels on the left are in the $[\bar{1}12]$ direction and the other set is in the $[1\bar{1}2]$ direction.}
	\label{Fig0}
\end{figure}

The ``$16\times2$'' reconstruction consists of a periodic channel structure which was identified by Ampo \textit{et al.} using low-energy electron diffraction (LEED) \cite{Ampo} with the channels parallel either to the $[\bar{1}12]$ or $[1\bar{1}2]$ directions \cite{Ichikawa,An,Stekolnikov,Kang}. Thus two different single domains can form as well as a combination of both (a double domain). The channels have been observed both by LEED and STM \cite{An,Ishikawa,Yamada} and measurements have shown a channel width of $2.5\,\,\si{nm}$ and a height of $0.19\,\,\si{nm}$. STM images of the reconstruction characteristically show `pairs of pentagons' on both levels of the periodic channel structure \cite{An}. A schematic diagram of a double domain reconstruction is shown in Fig. \ref{Fig0}. 
STM studies aimed at interpretation of the `pairs of pentagons' have been the main component of several experimental \cite{An,Sakamoto} and theoretical investigations \cite{Stekolnikov,Setvin}. Different models have been proposed to describe the reconstruction such as the adatom-tetramer-interstitial (ATI) model \cite{Stekolnikov}, the adatom-buckling model \cite{Sakamoto} and a multi-ringed structure \cite{Yamasaki}. The ATI model has been shown to contradict experimental evidence for step-edge states \cite{Setvin2010} and surface states close to the Fermi level \cite{Ferraro} and as such the atomic configuration remains controversial \cite{An,Stekolnikov,Sakamoto,Setvin,Yamasaki}. The same structure has been observed for the reconstructed Ge(110) surface \cite{Ichikawa,Mullet}.

The mechanism of single-domain formation has also been the subject of investigations. Initially, it was proposed that a single domain, the template for nanowires \cite{Hong}, is formed on surfaces that have been polished slightly off axis \cite{Ishikawa}, \textit{i.e.} surfaces that have been rotated along an axis in the surface plane such that the polished crystal face is at a small angle relative to the desired face. However, mechanistic studies have been limited in extent. In 2008, Yamada \textit{et al.} suggested that passing a current in the channel direction is key to generating macroscopic single domains, which occurs through the process of electromigration \cite{Yamada2}. However, this has been disputed by Sakamoto \textit{et al.} who proposed that the current direction does not influence single-domain formation \cite{Sakamoto}. More recently, Alguno \textit{et al.} showed that the current direction affects step structures on the Si(110) surface \cite{Alguno} but further experimental evidence is required to identify how a single domain is generated. 

The aim of this paper is to contribute to this debate by identifying the parameter that determines the direction in which the channels form in a single domain reconstruction. It is proposed that the domain orientation is determined by the direction of the fluctuating steps that are present at high-temperature. These are, in turn, determined by the off-axis polish direction. Such information is important for the development of nanowire templates as the direction of the nanowires can be known before the sample is introduced into vacuum. This proposal is supported by experimental evidence from X-ray diffraction, LEED and STM. Theoretical models are used to correctly identify the domains observed in reciprocal space and to support the observation from STM images that the surface is chiral \cite{An}. 

\section{Experimental method}
\begin{figure}[!hptb]
	\centering
	\includegraphics[width=\linewidth]{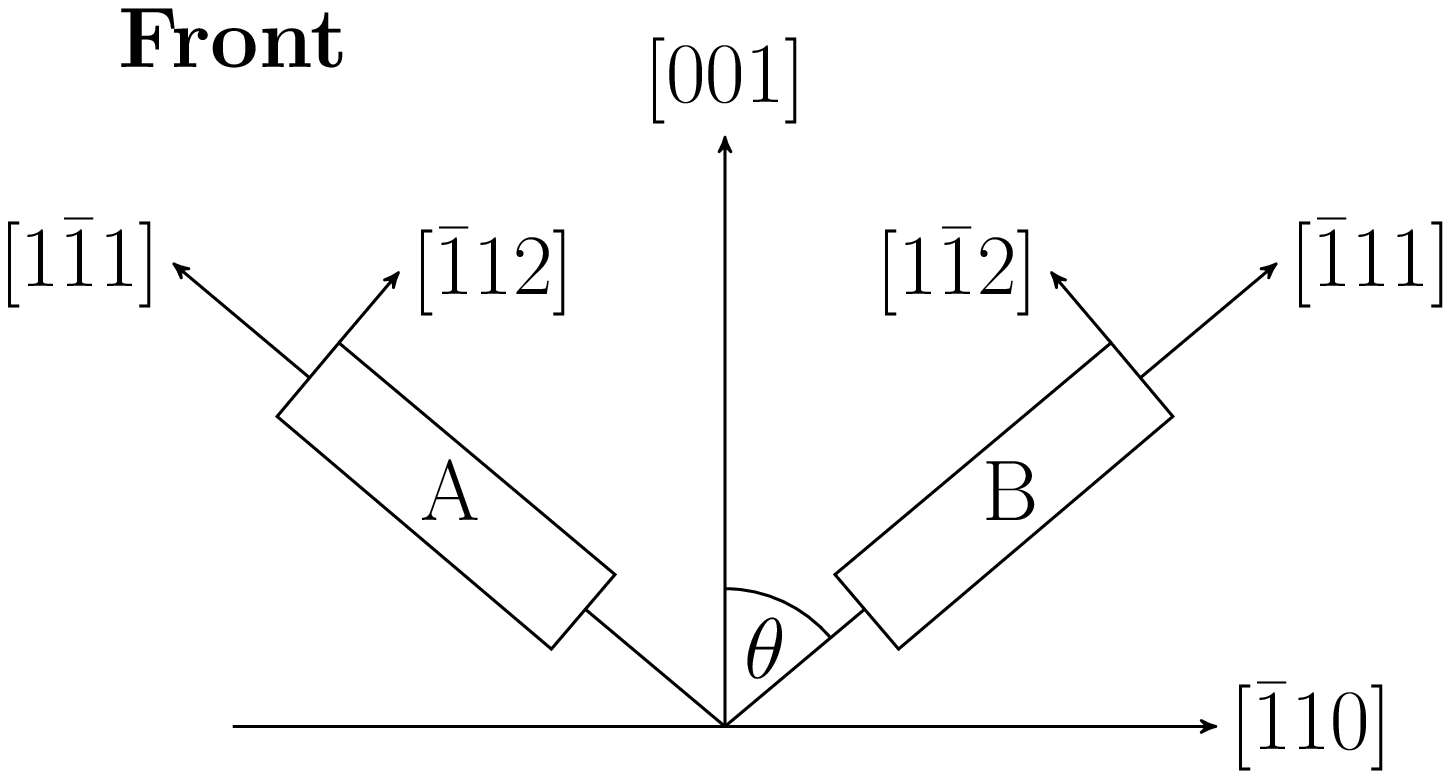}
	\includegraphics[width=\linewidth]{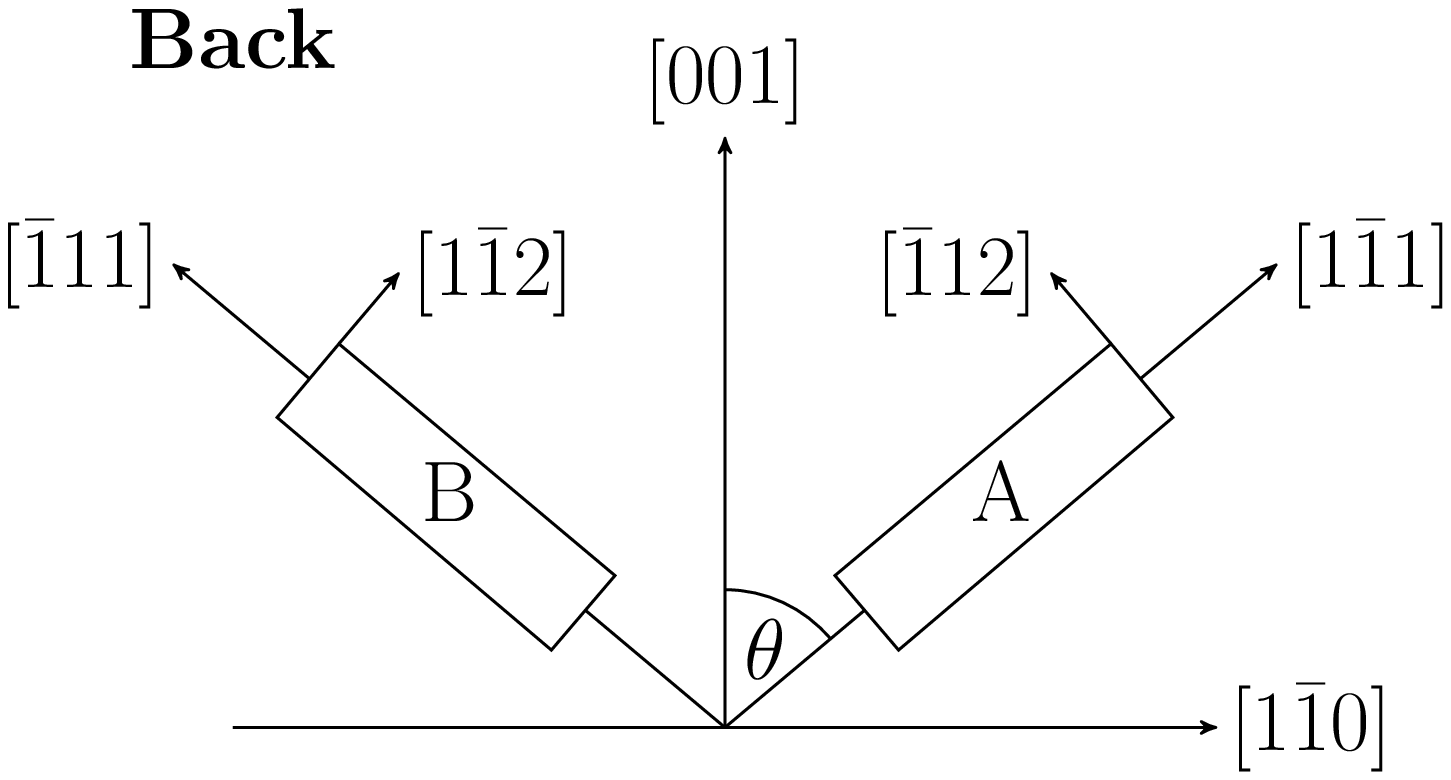}
	\caption{Diagrams of the front and back of two types of Si(110) samples, A and B, used in the experiments. The top figure defines the front of the samples and the bottom defines the back of the samples. The angle, $\theta$, indicated on the diagram is $54.7\,\si{\degree}$.}
	\label{Fig1}
\end{figure}

Phosphorus-doped silicon wafers (resistivity of $4-6\,\,\Omega\, \text{cm}$) supplied by PI-KEM Ltd were cut to expose the (110) surface and were polished on both sides. The wafers were then cut into rectangular samples of $12\times2\times0.25\,\,\si{mm}$ with their short edges in either the $[\bar{1}12]$ or $[1\bar{1}2]$ directions. The front and back of the samples are defined in Fig. \ref{Fig1} in order to remove ambiguity when interpreting LEED patterns. The different crystal orientations were confirmed by X-ray diffraction using an Oxford Diffraction X'Calibur. This apparatus uses a molybdenum $K_{\alpha}$ source with a wavelength of $0.7077\,\,\si{nm}$ and an Atlas S1 CCD detector. The data were analyzed using Rigaku CrysAlisPro software. This showed that the (110) faces were polished off-axis by $\sim 0.3\,\si{\degree}$. 

In all cases, the ``$16\times2$'' reconstruction was generated by resistive heating with the current direction along the long axes of the samples. After cleaning in acetone and isopropanol, the samples were degassed for about five hours at $650\,\si{\degreeCelsius}$ in UHV ($10^{-9}\,\,\si{mbar}$ or better). Temperature measurements were obtained using a pyrometer. Surface contaminants, such as oxygen and carbon, were removed by flashing the surfaces to $1200\,\,\si{\degreeCelsius}$ for between 1 and $2\,\,\si{s}$, a process repeated several times with one minute intervals. After flashing, the samples were annealed at $\sim710\,\,\si{\degreeCelsius}$ for 15 minutes and then gradually cooled at a rate of $50\,\,\si{mA/min}$ \cite{Yamada,Sakamoto}. When the current was rapidly reduced to zero after the annealing period, the Si(110) $1\times1$ surface resulted. After surface reconstruction, LEED and STM were performed. LEED experiments were carried out using OCI Vacuum Engineering LEED optics at room-temperature. STM was conducted using a purpose-built room temperature UHV STM apparatus on the APE-LE beamline at the Elettra synchrotron \cite{Panaccione}. The images were obtained using the WSxM software \cite{Horcas}.
\section{Results and discussion}
\begin{figure}[!hptb]
	\centering
	\includegraphics[width=0.45\linewidth]{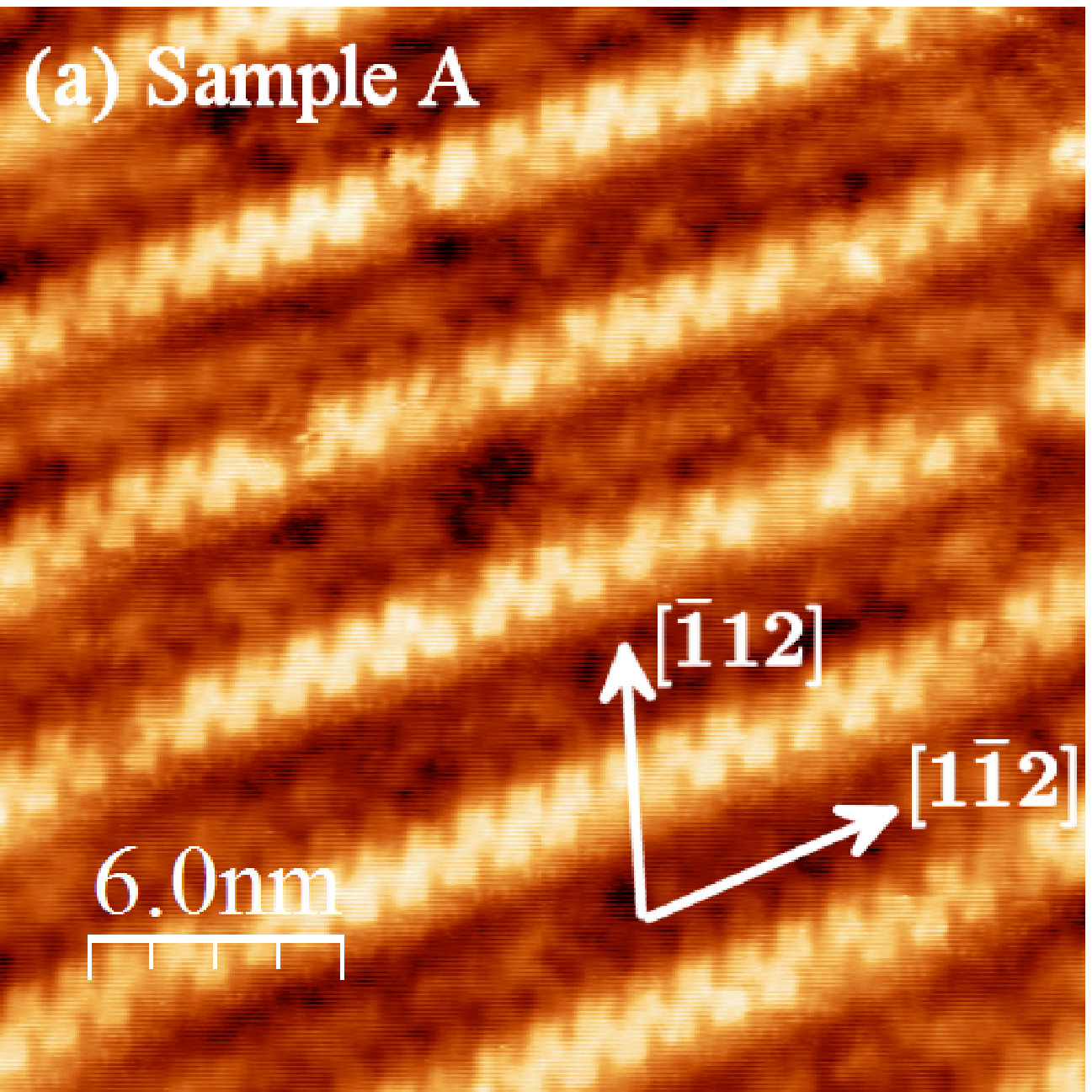}
	\includegraphics[width=0.45\linewidth]{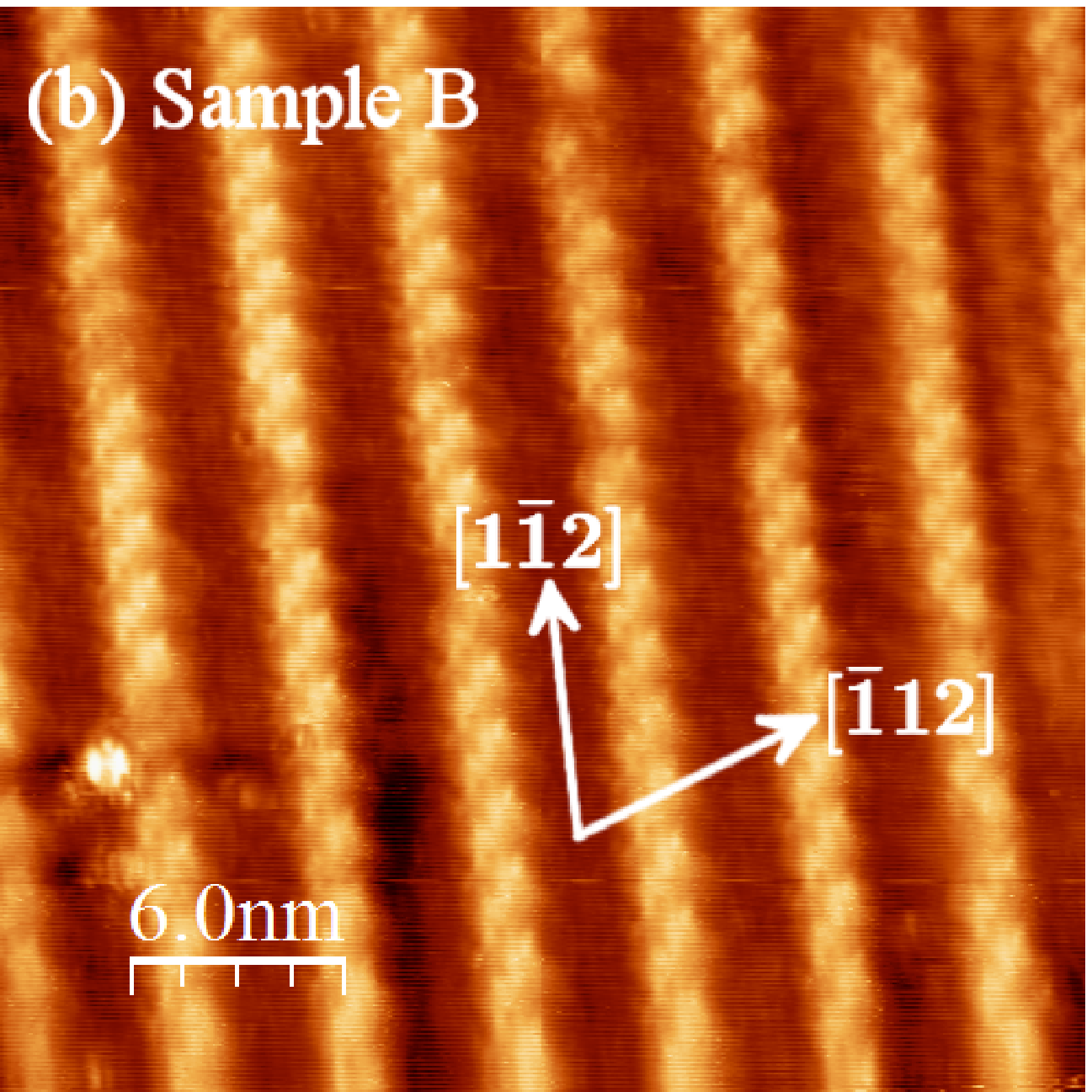}
	\caption{STM images of (a) Sample type A and (b) Sample type B showing the Si(110)``$16\times2$'' reconstruction. Both images show channels in the $[1\bar{1}2]$ direction and the current was applied approximately in the horizontal $[1\bar{1}1]$ and $[\bar{1}11]$ directions, respectively. The images were taken at a bias voltage of $1.6\,\,\si{V}$ and a tunneling current of $0.9\,\,\si{nA}$.}
	\label{Fig2}
\end{figure}

\begin{figure}[!hptb]
	\centering
	\includegraphics[width=\linewidth]{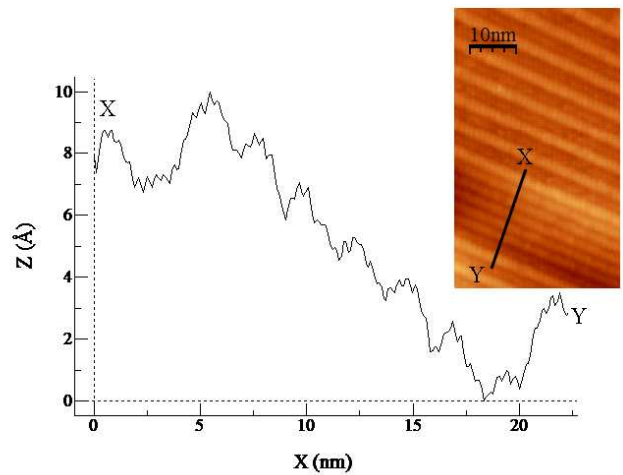}
	\caption{Line profile taken between X and Y on the inserted STM image of Sample type A showing the Si(110)``$16\times2$'' reconstruction. This shows step bunching present on the surface with 7 steps with a separation of $\sim 2.5\,\,\si{nm}$ and a height difference of $\sim 0.15\,\,\si{nm}$ \cite{Yamamoto1994I}.}
	\label{Fig3}
\end{figure}

STM images of Sample types A and B are shown in Fig. \ref{Fig2}(a) and \ref{Fig2}(b), respectively. 
The samples were mounted horizontally such that the short axis was aligned approximately with the vertical direction in Figs. \ref{Fig2}(a) and (b). Before acquisition, the STM tip was positioned centrally and a single-domain structure was observed over at least $25\times 25$ and $30\times30\,\,\si{nm}$ on Sample types A and B, respectively. A larger scale STM image of $100\times100\,\,\si{nm}$ is presented in Fig. 1 of the supplemental material \cite{Supp}. Figure \ref{Fig2}(a) shows the channels of Sample type A at an angle of $70\,\si{\degree}$ to the short edge of the sample. This means that these channels are in the $[1\bar{1}2]$ direction, because the short axis of Sample type A is in the $[\bar{1}12]$ direction which lies at $70\,\si{\degree}$ to the $[1\bar{1}2]$ direction. Figure \ref{Fig2}(b) shows that the channels of Sample type B are rotated by about $4\,\si{\degree}$ from the vertical, due to misalignment of the sample holder. Nevertheless, the channels are parallel to the short axis of the crystal in the $[1\bar{1}2]$ direction. STM measurements were made on both faces of the samples which showed that the channels lie in all cases in the $[1\bar{1}2]$ direction. Step-bunching of the fluctuating steps was also observed on these surfaces using STM, and Fig. \ref{Fig3} shows a line profile of a facetted surface of Sample type A after reconstruction. The STM image in Fig. \ref{Fig3} clearly shows that the channels are parallel to the edge direction of the bunched steps. Figure 2 of the supplemental material shows step bunching over a larger scale \cite{Supp}.

The channel direction of the reconstruction on Sample type B was found to be perpendicular to the heating current direction (which was parallel to the long axis) indicating that electromigration is not the driving force for the domain formation. Furthermore, the fact that Sample type A also has channels in the $[1\bar{1}2]$ direction indicates that there is some property of the surfaces common to both samples that determines the channel direction. 

\begin{figure}[!hptb]
	\centering
	\includegraphics[width=0.45\linewidth]{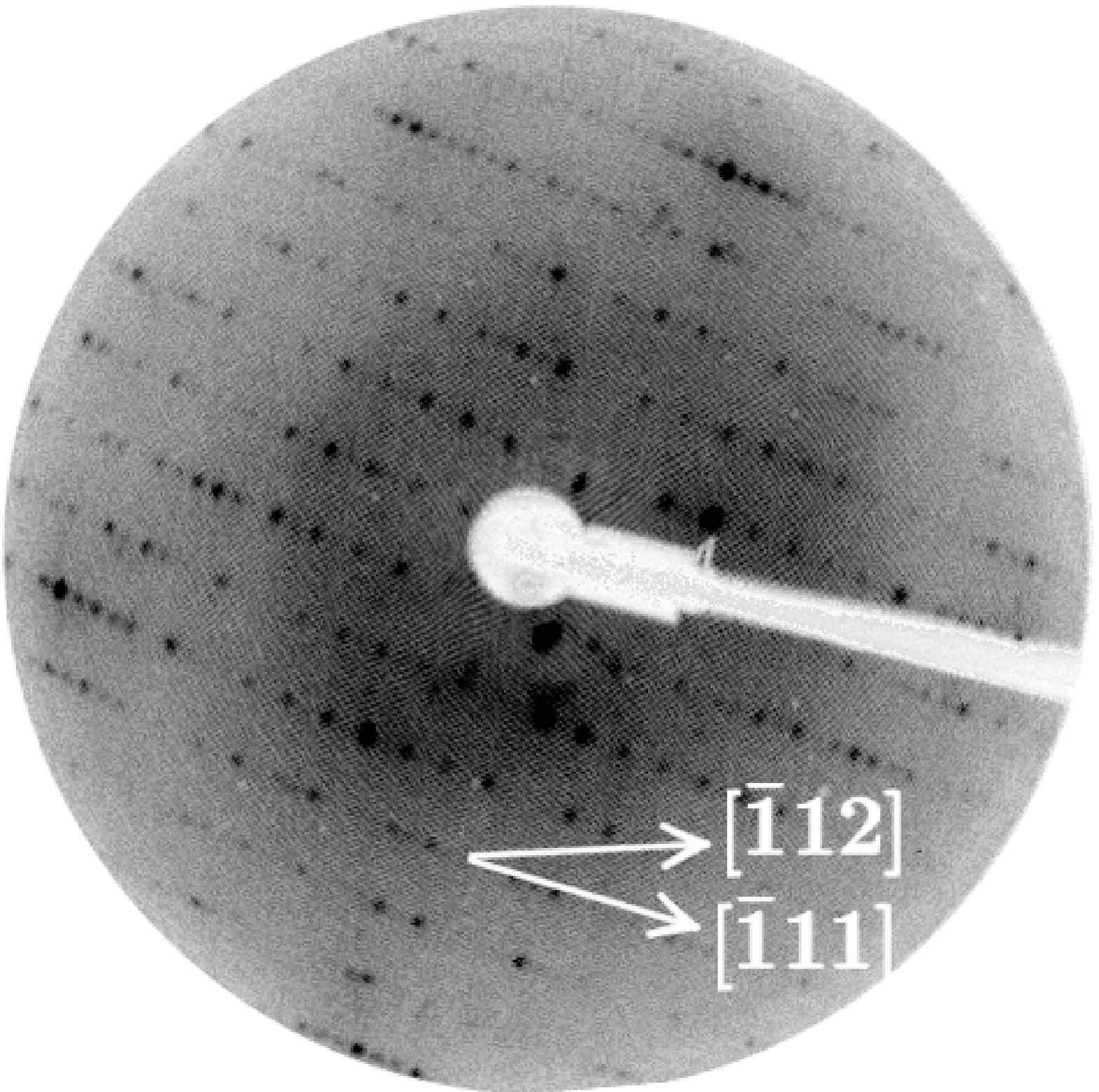}	
	\includegraphics[width=0.45\linewidth]{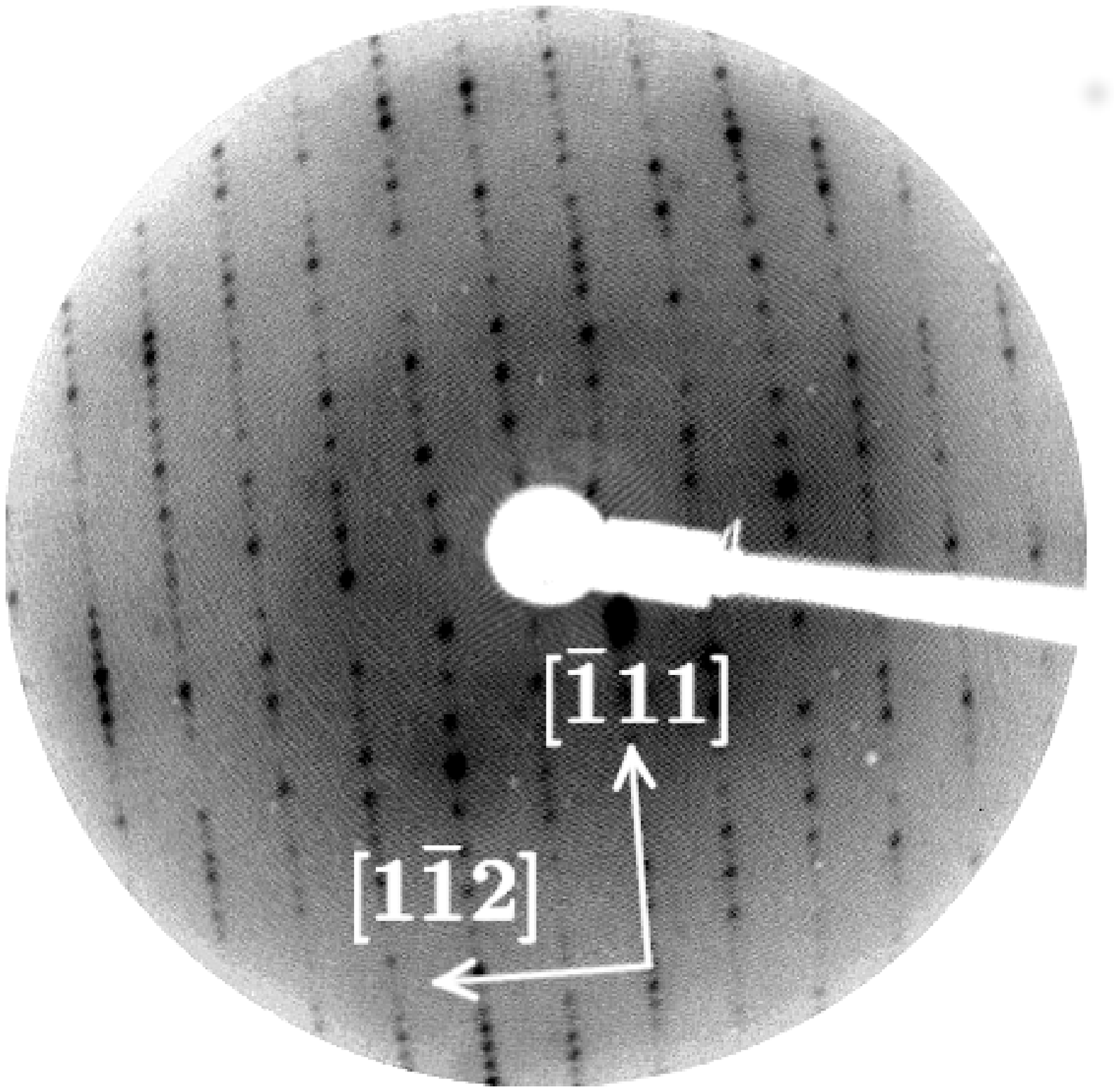}
	\newline \hspace*{0.2cm} (a) Sample type A \hspace*{1cm} (b) Sample type B
	\newline \includegraphics[width=0.45\linewidth]{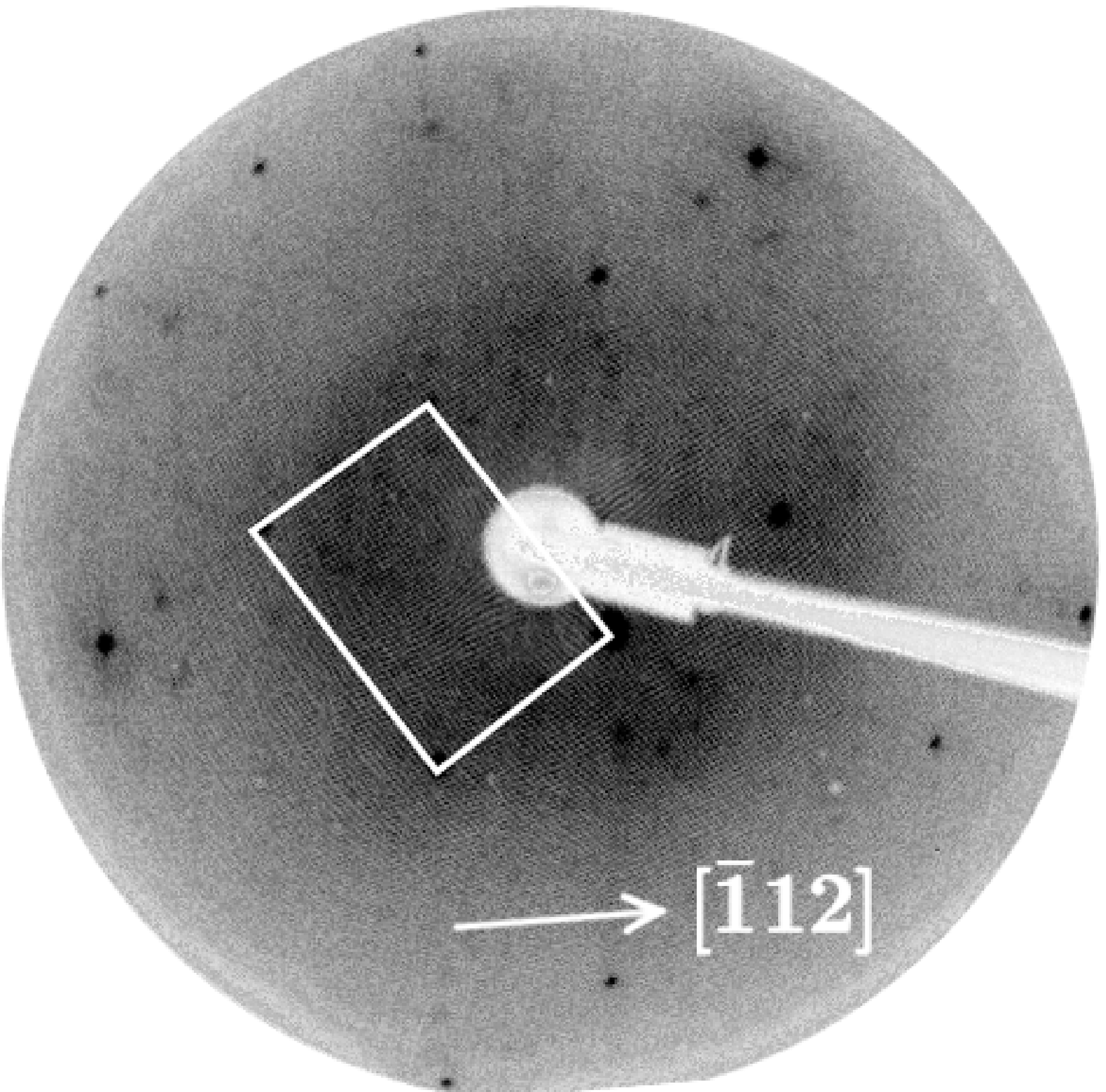}
	\includegraphics[width=0.45\linewidth]{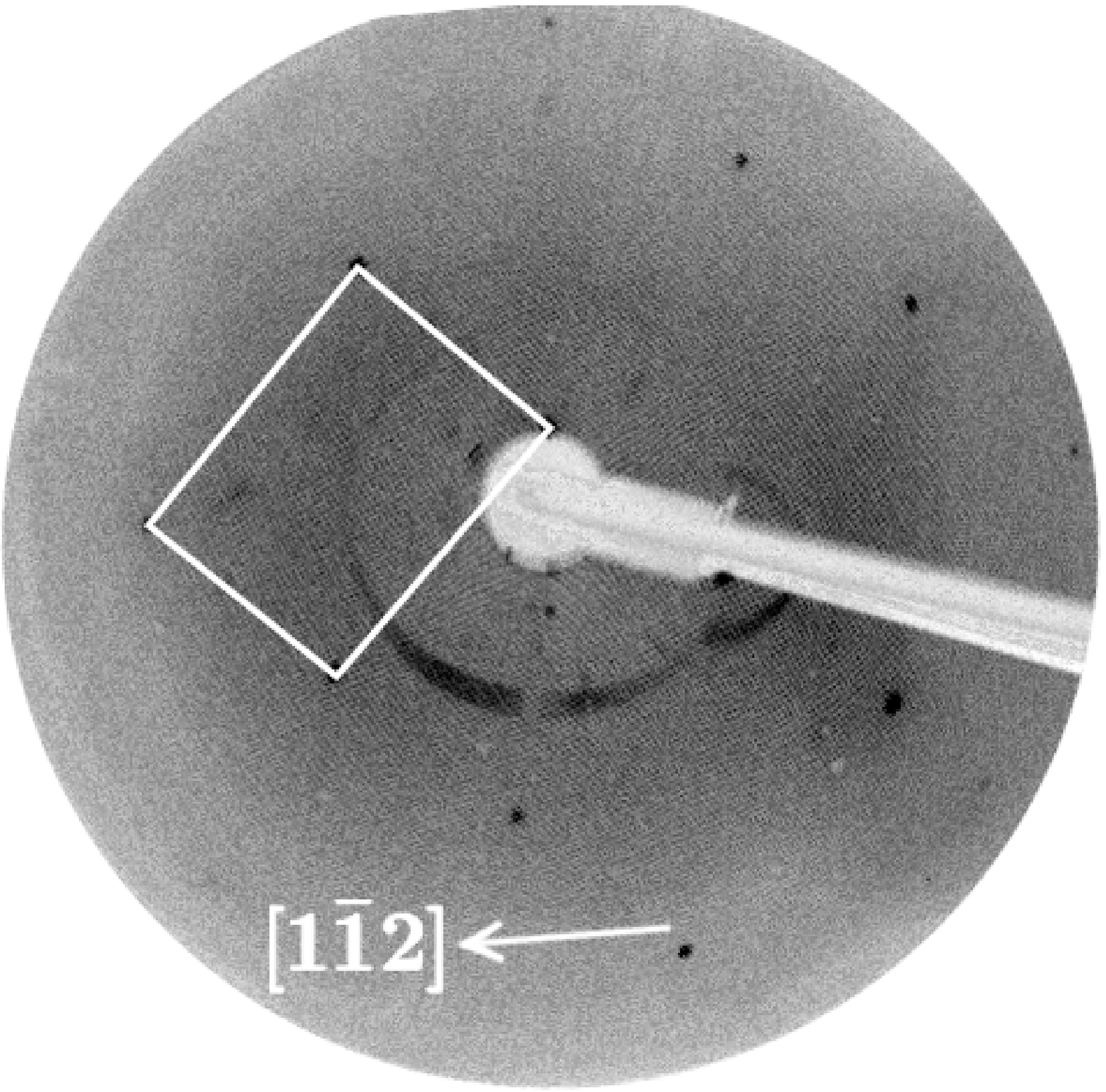}
	\newline \hspace*{-0.4cm} (c) Sample type A \hspace*{1cm} (d) Sample type B
	\caption{(a) and (b) are the LEED images of the ``$16\times2$'' reconstruction for Si(110) surfaces with the sample short axes along $[\bar{1}12]$ and $[1\bar{1}2]$, respectively. Both images show the fractional-order spots in the $[\bar{1}11]$ direction. (c) and (d) are the 1$\times$1 surface LEED patterns for the samples in (a) and (b), respectively. All directions shown are real-space directions. The unit cells of the LEED patterns for the 1$\times$1 surfaces are shown by white rectangles. (a)-(c) were obtained at a primary energy of $67\,\,\si{eV}$ whereas (d) was obtained at $56\,\,\si{eV}$.}
	\label{Fig4}
\end{figure}

Figures \ref{Fig4}(a) and \ref{Fig4}(b) show LEED patterns for the Si(110)``$16\times2$'' reconstruction for Sample types A and B, respectively. The fractional-order spots are predicted to be in the real-space $[\bar{1}11]$ direction \cite{Ampo} for both surfaces because the channels are parallel to the $[1\bar{1}2]$ direction which is at $90\,\si{\degree}$ to the axes connecting the spots. However, the $[\bar{1}12]$ and $[1\bar{1}2]$ directions in Figs. \ref{Fig4}(a) and \ref{Fig4}(b) respectively, are not perfectly horizontal due to small misalignments of the sample holder. 

Both LEED images show a single domain indicating that where the LEED pattern was obtained the surface contains channels in only one direction. The long-range order was determined by moving the surface over the LEED electron beam and searching for changes in the pattern. The only positions on the samples where the patterns shown in Figs. \ref{Fig4}(a) and \ref{Fig4}(b) were not observed was at the ends of the samples close to the tantalum retaining clips. By moving the pyrometer over the surface, the temperature at these positions was observed to be lower than the rest of the sample. Hence uniform heating is important for the formation of a single domain consistent with the findings of Sakamoto \textit{et al.} \cite{Sakamoto}. 

To determine how the annealing time and cooling rate affect the channel directions, these were varied between five and 30 minutes and the effect on the LEED patterns monitored. The patterns generated were always found to be consistent with Figs. \ref{Fig4}(a) and \ref{Fig4}(b). This implies various annealing times and cooling rates have little affect on the ``$16\times2$'' reconstruction. These results add weight to the suggestion that a property of the surface determines the channel direction. 

The 1$\times1$ surface was produced when the current was quenched rapidly to zero after annealing. Figures \ref{Fig4}(c) and \ref{Fig4}(d) show the LEED patterns for the 1$\times1$ surface for Sample types A and B, respectively. The ideal $1\times1$ unit cells are rectangles, with long-to-short-side ratios of $\sqrt{2}$ (see Eq. \ref{Eq1}), and are linked by a rotation of $109.6\,\si{\degree}$. The LEED patterns for the ``$16\times2$'' and 1$\times$1 surfaces are connected by noting that the fractional-order spot directions in Figs. \ref{Fig4}(a) and \ref{Fig4}(b) are parallel to the diagonals of the unit cells marked in Figs. \ref{Fig4}(c) and \ref{Fig4}(d), respectively.

\begin{figure}[!htpb]
	\centering
	\includegraphics[width=\linewidth,trim={0 0 0 3cm}]{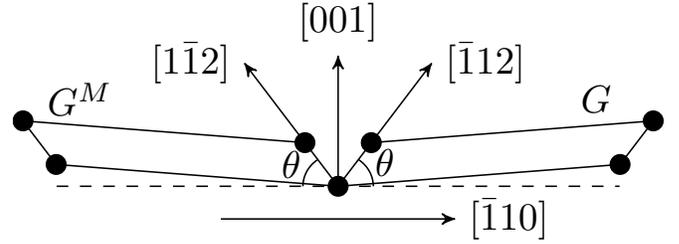} 
	\caption{Diagrams showing the unit cell parallelograms for the two possible domains of the ``$16\times2$'' reconstruction. For the front face, the $G$ and $G^M$ matrices represent the channels in the $[\bar{1}12]$ and $[1\bar{1}2]$ directions, respectively. The angle $\theta=54.7\,\,\si{\degree}$.}
\label{Fig5}
\end{figure}

Models were created to interpret the LEED patterns and to determine if the LEED and STM findings were consistent. The (110) bulk-terminated surface in real space is described by the lattice vectors 
\begin{equation}
\boldsymbol{a} = \frac{R}{\sqrt{2}}[\bar{1}10] \quad \text{and} \quad \boldsymbol{b} = R[001],
\label{Eq1}
\end{equation}
where $R$ is the lattice constant of silicon, $0.543\,\,\si{nm}$. These are related to the surface-lattice vectors, $\boldsymbol{r}_s$, through $\boldsymbol{r}_s = G\boldsymbol{r}$ where $G$ is a matrix specific to the surface under investigation \cite{Oura} and $\boldsymbol{r}$ is a column vector of $\boldsymbol{a}$ and $\boldsymbol{b}$. The $G$ matrix for the ``$16\times2$'' reconstruction is given by \cite{Bhattacharjee}
\begin{equation}
G = \begin{pmatrix} 2 & 2 \\ 17 & 1 \end{pmatrix}. \label{M1}
\end{equation}
An \textit{et al.} showed that the two possible ``$16\times2$'' domains are chiral \cite{An}. Thus from Fig. \ref{Fig1}, they are related by a mirror plane along the $[001]$ axis. Applying this to the $G$ matrix produces
\begin{equation}
G^{M} = \begin{pmatrix}
-2 & 2 \\ -17 & 1
\end{pmatrix}, \label{M2}
\end{equation}
representing the second possible domain orientation. Using these two matrices, the reconstructed surface-lattice vectors were calculated. Figure \ref{Fig5} shows the unit cells of the two domains of the Si(110)``$16\times2$'' reconstruction. The angle $\theta$ between the horizontal and the $[\bar{1}12]$ direction in Fig. \ref{Fig5} is equal to $54.7\,\si{\degree}$. Hence the matrices $G$ and $G^M$ represent the channels in the $[\bar{1}12]$ and $[1\bar{1}2]$ directions, respectively.

\begin{figure}[!hptb]
	\centering
	\includegraphics[width=0.45\linewidth]{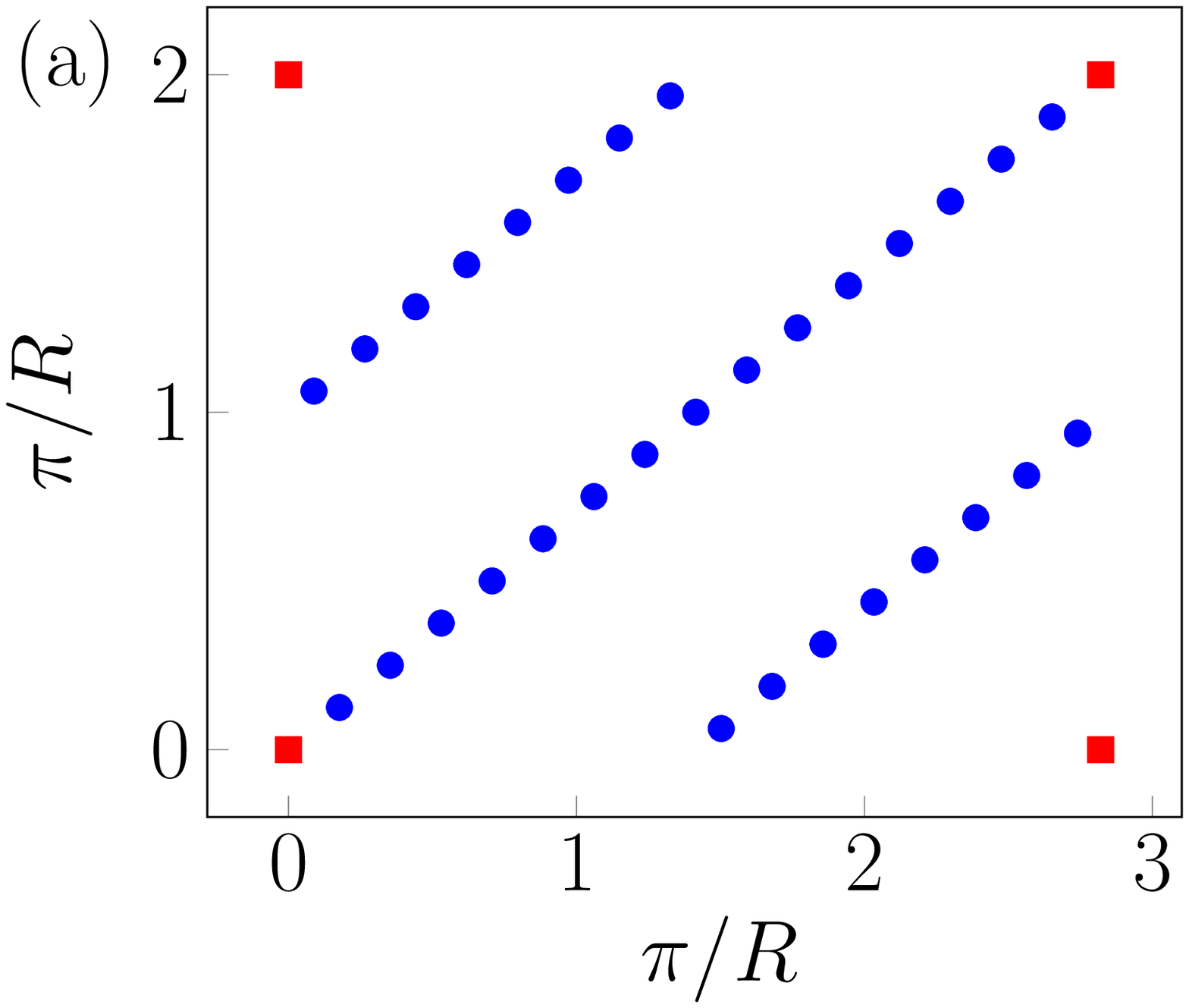}
	\includegraphics[width=0.45\linewidth]{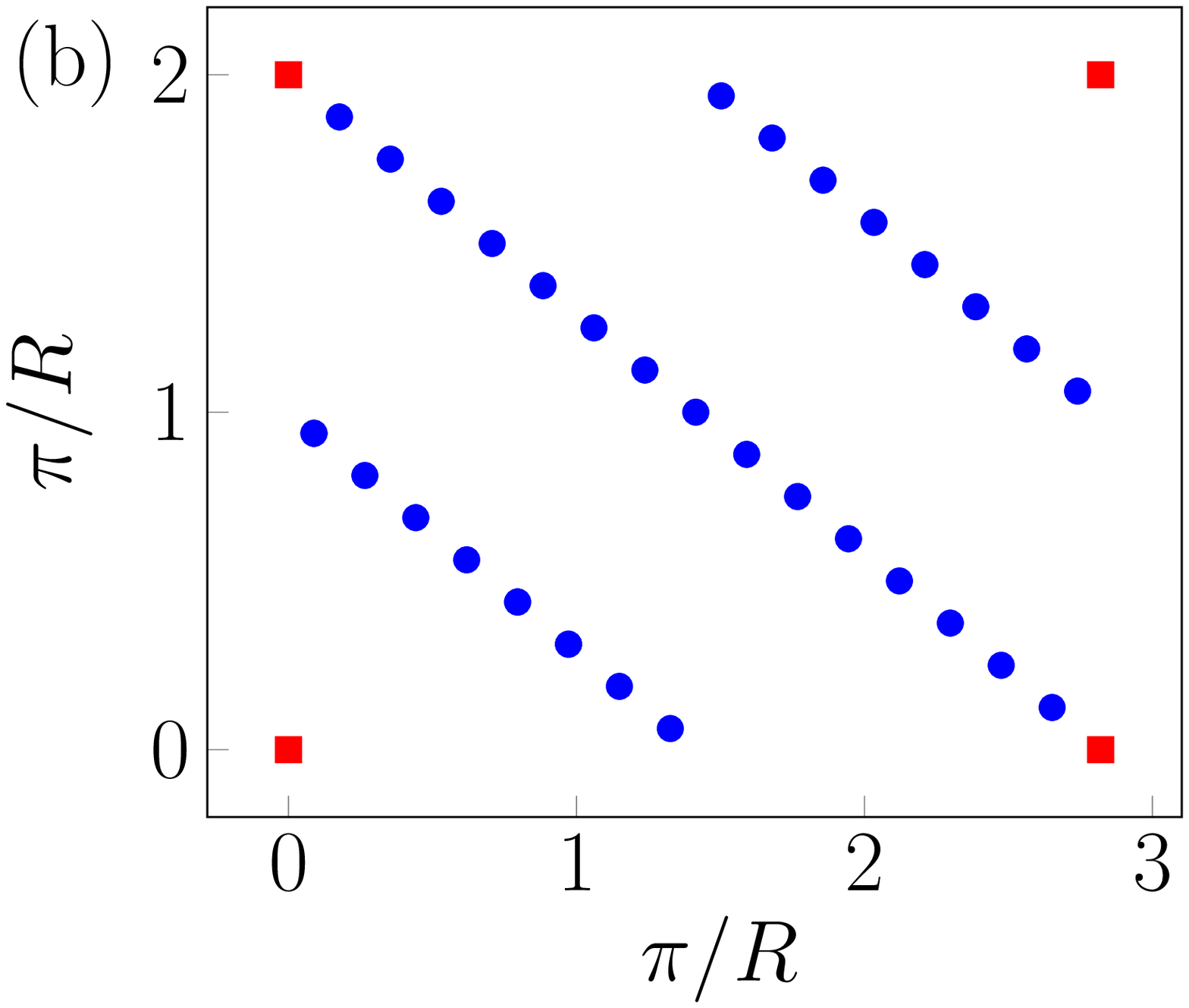}
	\caption{(a) and (b) are the reciprocal-space lattices of the two possible domains for the ``$16\times2$'' reconstruction calculated using $G^M$ and $G$, respectively; $R$ is the lattice constant of silicon. The blue circles were generated using $\boldsymbol{r}^*_s$ and the red squares were generated using $\boldsymbol{r}^*$ which correspond to the reciprocal lattice vectors of the reconstructed and bulk-terminated surfaces, respectively.}
	\label{Fig6}
\end{figure}

To relate the $G$ matrices and the LEED patterns the reconstructed reciprocal-lattice vectors, $\boldsymbol{r}^*_s$, were calculated using $\boldsymbol{r}_s^* = G^*\boldsymbol{r}^*$, where $G^* = \left(G^{-1}\right)^T$ \cite{Oura} and $\boldsymbol{r}^*$ is a column vector of $\boldsymbol{a}^*$ and $\boldsymbol{b}^*$ which are the bulk-terminated reciprocal-lattice vectors. Figures \ref{Fig6}(a) and \ref{Fig6}(b) show the reconstructed reciprocal lattices calculated using $G^M$ and $G$ respectively, with the fractional-order spots shown as filled blue circles and the bulk-terminated spots shown as red squares. Thus Figs. \ref{Fig6}(a) and \ref{Fig6}(b) represent channels in the $[1\bar{1}2]$ and $[\bar{1}12]$ directions respectively, for the front face. These lattices confirm the Si(110)``$16\times2$'' reconstruction is chiral. This is because neither reciprocal-space lattice contains a mirror plane which indicates that the corresponding real-space reconstructions do not contain any mirror planes. 

In order to interpret the LEED patterns using the predicted reciprocal lattices, the front and back faces, as defined in Fig. \ref{Fig1}, must be distinguished. This is important for identification of the channel directions using LEED patterns. Previous work of Yamada \textit{et al.} used the LEED patterns of the 1$\times$1 and ``$16\times2$'' reconstructions to show the two different domains \cite{Yamada}. However, ambiguities in real-space directions caused difficulties in determining the domain orientation. 

\begin{figure}[!hptb]
	\centering
	\includegraphics[width=0.45\linewidth]{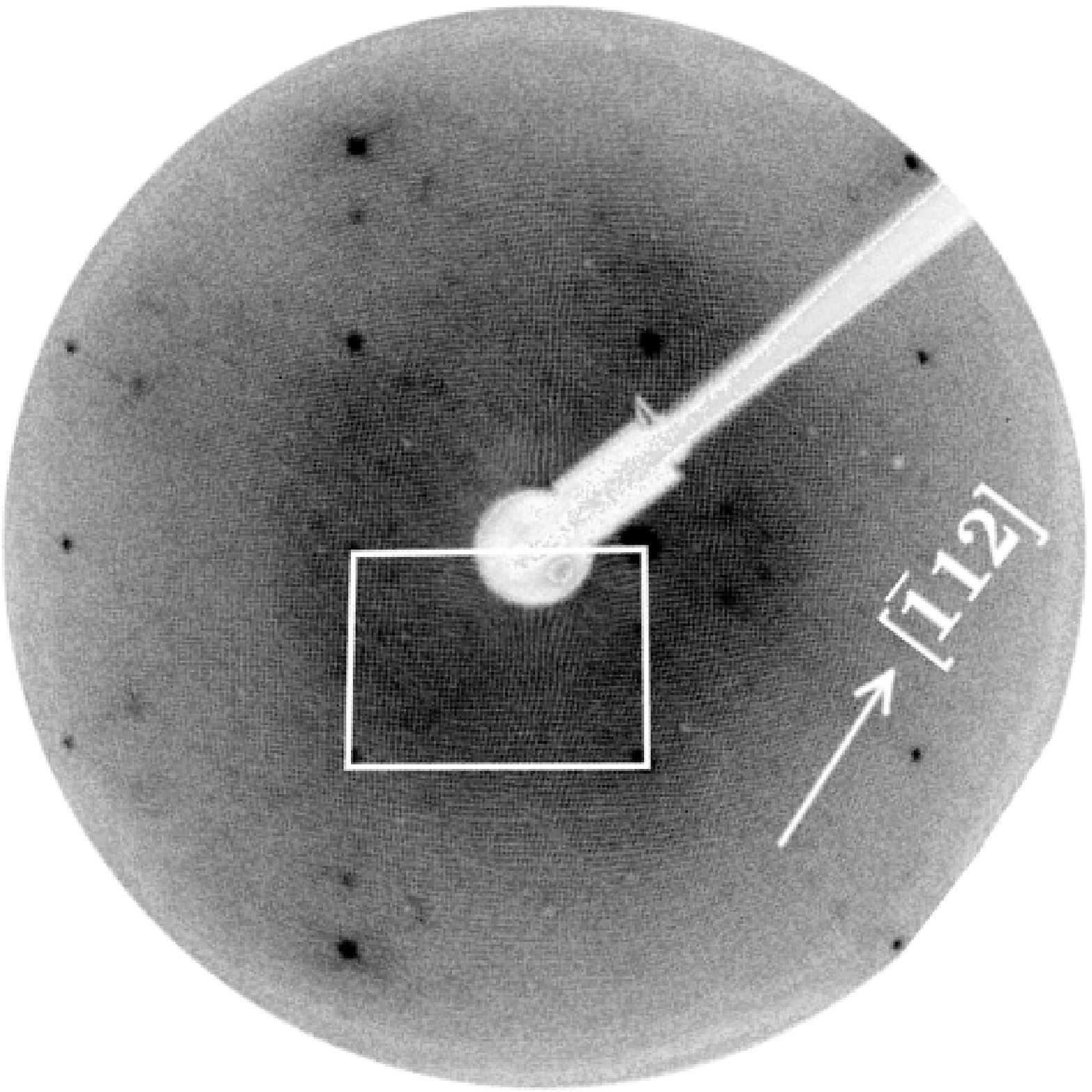}
	\includegraphics[width=0.45\linewidth]{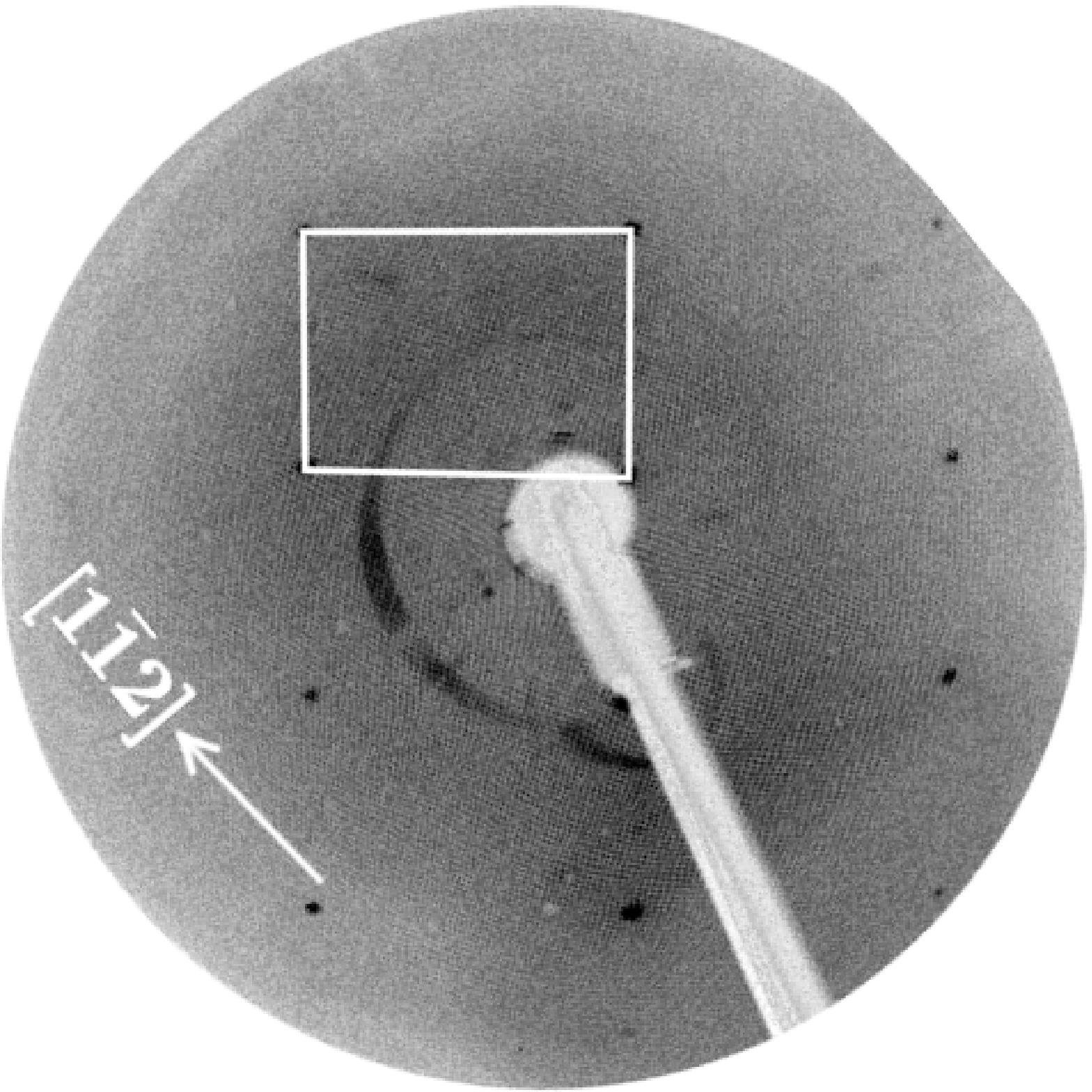}
	\newline \hspace*{-0.4cm} (a) Sample type A \hspace*{1.3cm} (b) Sample type B
	\caption{1$\times$1 LEED patterns from Fig. \ref{Fig4}(c) and (d) rotated through an angle of $+55\,\si{\degree}$ for (a) and $-55\,\si{\degree}$ for (b).}
	\label{Fig7}
\end{figure}

The front or back of the Si(110) crystal is experimentally ascertained by using the 1$\times$1 LEED pattern and first rotating it through either $+55\,\si{\degree}$ or $-55\,\si{\degree}$ such that the pattern observed is in the same orientation as the red squares in Figs. \ref{Fig6}(a) and \ref{Fig6}(b). Then by comparing the crystal directions on the 1$\times$1 LEED patterns with those in Fig. \ref{Fig1}, the front or back can be identified. Using this procedure for the LEED patterns in Figs. \ref{Fig4}(c) and \ref{Fig4}(d) indicates that these show the front faces because when rotated, the $[\bar{1}12]$ direction in Fig. \ref{Fig4}(c) and the $[1\bar{1}2]$ direction in Fig. \ref{Fig4}(d) align with the front faces in Fig. \ref{Fig1}. The rotated $1\times1$ LEED patterns are shown in Figs. \ref{Fig7}(a) and \ref{Fig7}(b).

\begin{figure}
	\centering
	\includegraphics[width=0.45\linewidth]{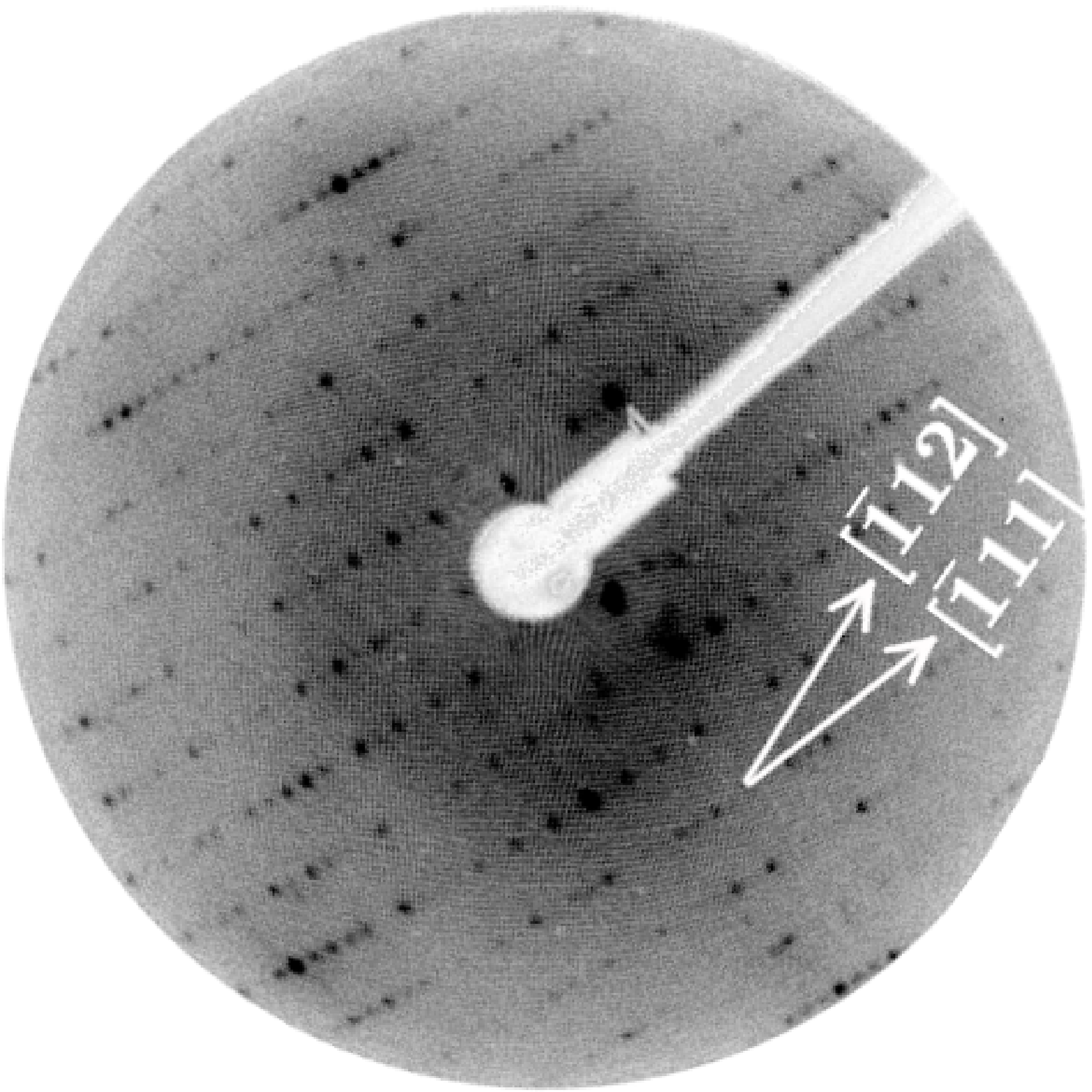}
	\includegraphics[width=0.45\linewidth]{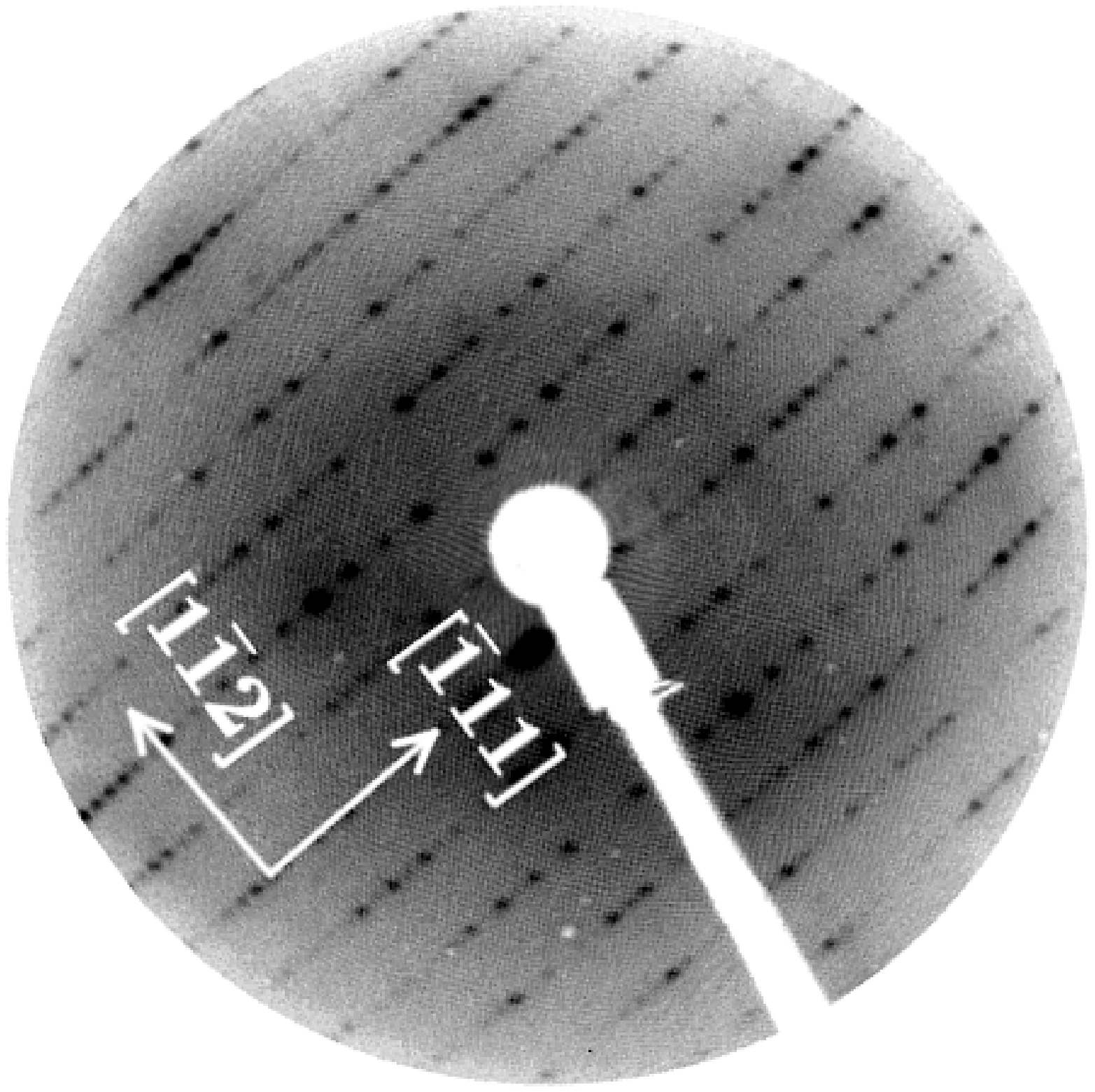}
	\newline \hspace*{0cm} (a) Sample type A \hspace*{1.3cm} (b) Sample type B
	\newline \hspace*{0.6cm}\includegraphics[width=0.45\linewidth]{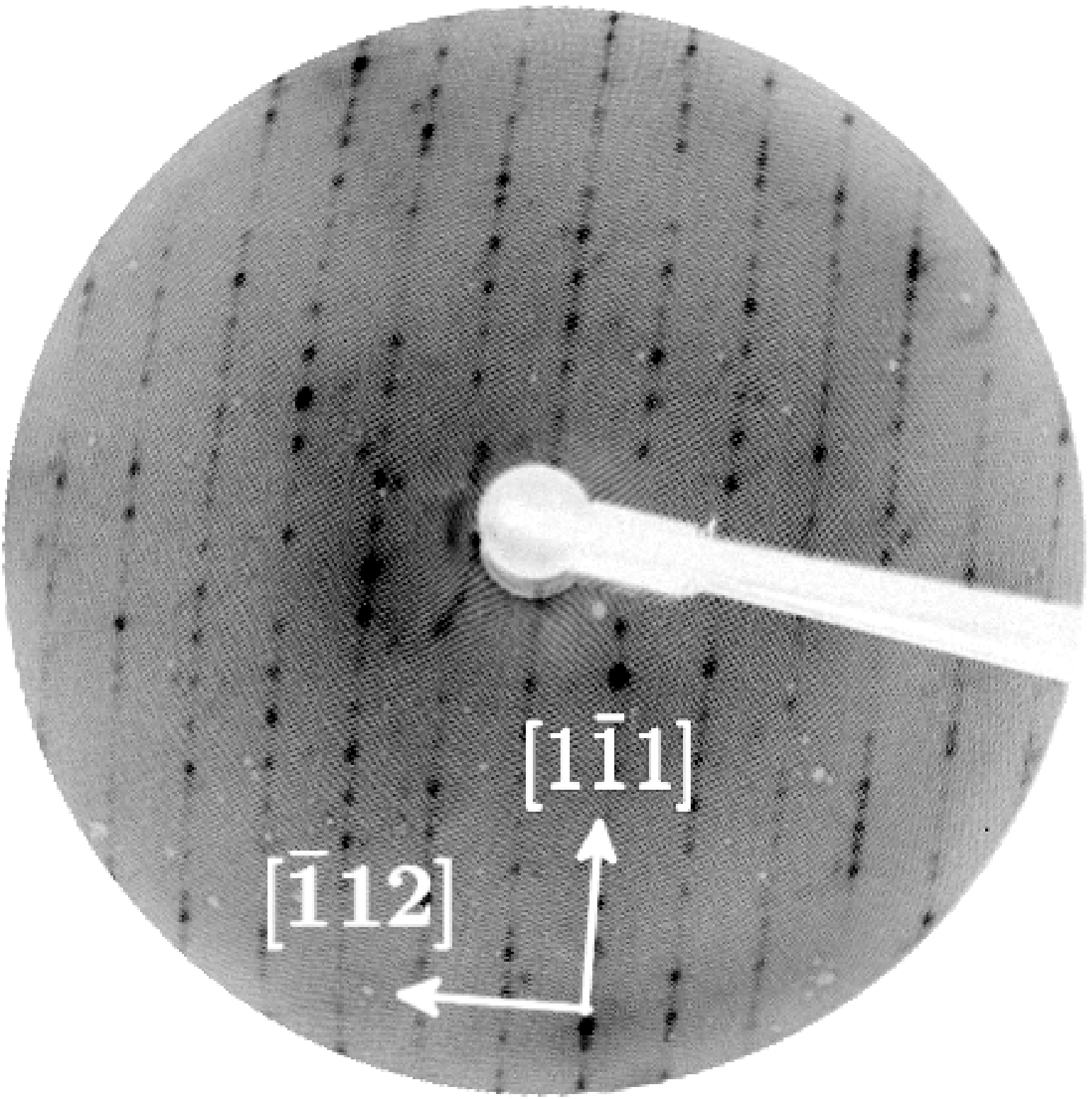}
	\newline \hspace*{-0.7cm}(c) Sample type A
	\caption{``16$\times2$'' LEED patterns from Fig. \ref{Fig4}(a) and (b) rotated through an angle of $+55\,\si{\degree}$ for (a) and $-55\,\si{\degree}$ for (b). (c) is the LEED pattern of the Si(110)``$16\times2$'' reconstruction for Sample type A with an off-axis polish angle changed to $0.5\,\si{\degree}$ about the $[\bar{1}12]$ rotation axis. The $1\times1$ LEED pattern for the same sample in (c) reveals it to be the back face.}
	\label{Fig8}
\end{figure}

The channel directions are obtained by observing which diagonal of the unit cell in Figs. \ref{Fig7}(a) and \ref{Fig7}(b) the fractional-order spots lie along and then comparing that direction with the blue circles in Fig. \ref{Fig6}(a) and \ref{Fig6}(b). It can be seen that the rotated ``$16\times2$'' LEED patterns in Figs. \ref{Fig8}(a) and \ref{Fig8}(b) match the lattice in Fig. \ref{Fig6}(a). Thus the channels are in the $[1\bar{1}2]$ direction. This confirms the STM results. If a sample is identified as the back face the channel directions are swapped in Fig. \ref{Fig5} such that $G$ represents channels in the $[1\bar{1}2]$ direction and $G^M$ represents channels in the $[\bar{1}12]$ direction. Thus the reciprocal space lattices in Figs. \ref{Fig6}(a) and \ref{Fig6}(b) would correspond to channels in the $[\bar{1}12]$ and $[1\bar{1}2]$ directions.

Analysis of the LEED and STM images has thus shown that the heating current direction, or more explicitly electromigration, is not pivotal in single domain formation or for determining the domain orientation (\textit{i.e.} the direction of the channels) of the Si(110)``$16\times2$'' reconstruction. Variations in annealing time and cooling rate also had minimal influence on the reconstruction orientation. Furthermore, this experiment was repeated on 19 different samples both at Daresbury Laboratory and at the Elettra synchrotron (Trieste) which all showed the same domain orientation. Therefore, it is unlikely any random variable is determining the domain orientation.

\begin{figure}[!hptb]
	\centering
	\includegraphics[width=\linewidth]{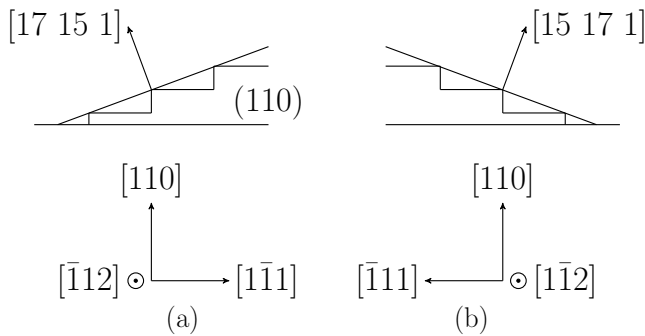}
	\caption{Crystal direction diagrams showing the orientations of (a) the (17 15 1) and (b) the (15 17 1) planes relative to the (110) plane. The angle between the [17 15 1] and [110] directions is $4.3\,\si{\degree}$. The same applies for the [15 17 1] and the $[110]$ directions.}
	\label{Fig9}
\end{figure}

Ishikawa \textit{et al.} reported \cite{Ishikawa} that a single domain is formed on reconstruction of a Si(110) surface that is polished slightly off-axis. The samples in this experiment were initially polished off-axis by $\sim0.3\,\si{\degree}$ and the same domain orientation was observed for all samples after reconstruction. It is therefore proposed that the single domain observed is a result of the direction of the monoatomic steps caused by the off-axis polish direction. If a Si(110) surface is polished such that the off-axis rotation is along the $[\bar{1}12]$ or $[1\bar{1}2]$ directions, this can result in vicinal (17 15 1) or (15 17 1) surfaces respectively (see Fig. \ref{Fig9}(a) and \ref{Fig9}(b)). 
Then after surface reconstruction the domain which forms depends on the direction of the monoatomic steps. This hypothesis was tested by re-polishing Sample type A wafers, to have an off-axis angle of $0.5\,\si{\degree}$ along the $[\bar{1}12]$ direction so as to favor the vicinal (17 15 1) surface, see Fig. \ref{Fig9}(a). Evidence presented in Figs. \ref{Fig8}(a) and \ref{Fig8}(b) suggests that the (15 17 1) plane (Fig. \ref{Fig9}(b)) is favored for the original samples giving rise to the $[1\bar{1}2]$ channel directions. 

The LEED pattern of the ``$16\times2$'' reconstruction of a re-polished sample is shown in Fig. \ref{Fig8}(c). The reconstruction has been conducted on five re-polished samples and all showed spots along the vertical direction, parallel to the long edge of the sample, which indicates that the channels are along the short axis of the sample. Thus the channels are along the $[\bar{1}12]$ direction. The channel direction was also confirmed by comparing the reciprocal-space lattices of the 1$\times1$ and ``$16\times2$'' surfaces. Comparison of Figs. \ref{Fig8}(a) and \ref{Fig8}(c) shows that the re-polished Sample type A has a different channel direction on reconstruction. This shows that the fluctuating steps influence the domain orientation observed after surface reconstruction, because the surface was polished to favor the (17 15 1) surface. If the off-axis rotation lies in a general direction in the (110) plane, that direction has components along the $[\bar{1}12]$ and $[1\bar{1}2]$ directions. Thus it is proposed the component with the larger coefficient will prevail in the reconstruction process \cite{Alguno}. However, a single domain can only form under the correct reconstruction conditions, \textit{i.e.} an annealing temperature of $700\,\si{\degreeCelsius}$ and a gradual cooling of the surface to room temperature. 

This domain formation mechanism is important for nanowire templates. Polishing a surface along either the $[\bar{1}12]$ or $[1\bar{1}2]$ directions, as shown in Figs. \ref{Fig9}(a) and \ref{Fig9}(b), will produce a single domain because monoatomic steps in the other direction are not present. This allows for a reliable method of producing large scale nanowires by using the ``16$\times$2'' reconstruction as a template. Furthermore, the mechanism allows for other methods of heating the surface, such as electron-beam heating, to be used to produce a single domain reconstruction if the temperature over the surface is uniform.

\section{Conclusions}
In summary, Si(110) surfaces that are polished off-axis can reconstruct to exhibit long-range order independent of the current direction used to heat them. It has been shown that electromigration, annealing time and cooling rate do not affect the domain orientation of the Si(110)``$16\times2$'' reconstruction. However, influencing the vicinal step direction, as a result of off-polishing a sample along the $[\bar{1}12]$ direction, has been shown to affect the domain orientation observed after reconstruction. A procedure has been outlined that allows for identification of the correct channel direction using experimental LEED patterns and models of the reciprocal-space lattice. The chirality of the Si(110)``$16\times2$'' reconstruction has also been confirmed.
\newline
\begin{acknowledgements}
This work was supported by EPSRC (UK) under Grant number EP/M507969/1. The research leading to these results received funding from the European Community's Seventh Framework Programme (FP7/2007-2015) under grant agreement 288879. Funding was also received from ASTeC and the Cockcroft Institute (UK), and the U.S. National Science Foundation (Awards PHY-1505794 and PHY-1430519; EB, NC and TG). The authors personally acknowledge: J. A. Smerdon and D. Bowler for their many useful discussions. The data associated with the paper are openly available from The University of Manchester eScholar Data Repository: \url{dx.doi.org/10.15127/1.306088}
\end{acknowledgements}

\end{document}